# Designing a Kinetic Façade Using BB-BC Algorithm with a Focus on Enhancing Building Energy Efficiency

**Marzieh Soltani*,Arash Atashi**

1. Master of Architecture, Islamic Azad University, Science and Research,Tehran Branch, Zanjan, Iran, marzieh.soltani@gmail.com
2. Master's student in Structures engineering, Islamic Azad University, Bushehr Branch, Bushehr, Iran, arash.atashi@yahoo.com

## Abstract

In order to increase energy efficiency in buildings, optimizing the parameters of the facade form can be challenging due to the dynamic nature of solar radiation. One effective solution is the use of kinetic facades as a second skin, which can control energy consumption. This study proposes a parametric kinetic facade to increase building energy efficiency, along with a framework to optimize its form using the Bang-Big Crunch (BB-BC) optimization algorithm. The study involved modeling a two-story office building in Shiraz city and calculating the energy consumption resulting from building operation over a three-day period without considering the second skin of the facade. In the second stage of the study, the second skin was optimized for the same three-day interval and calculated as a parametric, static facade. In the last step, the parameters of the second skin were optimized for three one-day intervals, assuming the possibility of kinematic changes each day. The total energy consumption of building operation for the three days was then calculated and analyzed.The Python programming language was used to develop the optimization algorithm, while Rhino software, and Grasshopper, Ladybug, and Honeybee plugins were used for building modeling and simulation of light, energy, and weather parameters.The results of the study demonstrate the effectiveness of the proposed kinetic facade and the proper performance of the proposed algorithm for solving similar problems. The study found that the use of the second skin with kinetic function reduced energy consumption by 28%. Additionally, the results from the second and third stages of the study showed that the use of the second facade shell with kinematic function, compared to its function in static mode, reduced energy consumption by 4%.Overall, the study highlights the potential of kinetic facades to control energy consumption in buildings and the importance of considering dynamic factors such as solar radiation when optimizing building design.

## 1. Introduction

The building and construction sector is responsible for a significant portion of final energy consumption and energy-related emissions, making it a crucial area for reducing greenhouse gas emissions. According to the International Energy Agency (IEA), the building sector accounts for approximately 36% of final energy consumption and 39% of energy-related carbon dioxide emissions worldwide [1].To address this issue, researchers have explored various methods to reduce energy consumption in buildings. One approach is to optimize the building form to improve energy efficiency. Studies have shown that optimizing building forms can lead to energy savings of up to 14.9% to 29% [2].Another solution for reducing energy consumption is the use of kinetic facades. Kinetic facades are composed of moving elements that can adjust to environmental conditions, such as the position of the sun, wind, and temperature, in order to improve energy efficiency and thermal comfort [3]. The integration of kinetic facades in building design has been shown to significantly reduce energy consumption and improve the indoor environment [4].There are three categories of kinetic facades: deployable, dynamic, and embedded [5]. Deployable facades are temporary structures that can be easily assembled and disassembled. Dynamic facades are entities within a larger architecture with independent responses to the environment. Embedded facades are located in a fixed position within a larger architecture and can be directly measured and analyzed.Recent research has explored the potential of multi-purpose kinetic facades to improve energy efficiency and environmental adaptability. For example, researchers have used optimization algorithms, such as the BB-BC algorithm, to design and compare kinetic form lists and their energy efficiency with static facades [6]. The potential of multi-purpose kinetic facades to improve energy efficiency and environmental adaptability has been demonstrated in recent research. This study aims to propose an embedded kinetic facade as a solution for increasing building energy efficiency and reducing energy consumption. To achieve this, the BB-BC algorithm is used to optimize the parameters of the kinetic facade and compare it with a static facade. The goal is to determine the energy savings and thermal comfort improvements achieved by the embedded kinetic facade compared to a static facade. The research findings will contribute to the development of sustainable building design strategies and further promote the use of kinetic facades in the construction industry.

## 2. Methodology

This study proposes a kinetic facade to increase the energy efficiency of a building and optimize its parameters using the BB-BC algorithm within a specific timeframe. The study consists of three stages. In the first stage, a two-story open-plan office building was modeled using Rhino software and Grasshopper, Ladybug, and Honeybee plugins, as well as the Python programming language to simulate light, energy, and weather parameters. The energy consumption of the building was calculated for three consecutive days in August. In the second stage, a geometrically flexible parametric facade was designed to respond to the kinematic changes of a kinetic view. This facade was placed as a second shell on the main facade. Assuming no kinematic changes of the kinetic facade, the BB-BC optimization algorithm was used to optimize the facade parameters during the same period of time. Once an optimal form of the facade was fixed, the results were calculated and analyzed..In the third stage, the facade was assumed to have the ability to change kinematics and parameters controlling the form every day. The results from all three stages were compared to each other to determine the effectiveness of the kinetic facade in increasing energy efficiency and optimizing building parameters.

## 2-1- Tools Used

### 2-1-1- Python Programming Language

Python is a powerful yet easy-to-learn programming language that follows an object-oriented approach. It was designed by Guido van Rossum [7].

### 2-1-2- Grasshopper

Grasshopper is a visual programming language and environmental program that runs within the Rhino environment. It was designed by David Rutten and works by combining different parts and components in a sequential manner within the program environment [8].

### 2-1-3- Honeybee and Ladybug

Honeybee is a simulation application used for daylight and thermodynamic analysis. It creates daylight simulation results by using radiance and energy models through its linkage with EnergyPlus and OpenStudio in the Grasshopper environment. Ladybug, on the other hand, imports standard files containing EnergyPlus weather information into Grasshopper. This program offers various 2D and 3D climate graphics that aid in the decision-making process during the early stages of design. Ladybug also supports the assessment of early design options through solar radiation studies, visibility analysis, sunlight hours modeling, and more. Integration with visual programming environments allows for immediate feedback on design decisions and provides a high degree of customization [9].

## 2-2- Building Modeling

The building being analyzed is a two-story office with an open plan, situated in Shiraz, and its weather data is sourced from the Shahid Dastghib airport station. The facade being studied is on the southern side of the building, and materials selection for the analysis was based on the ASHRAE 189.1-2009 standard. A three-dimensional view of the building can be seen in Figure 1, while Table 1 presents the various parameters of the building.

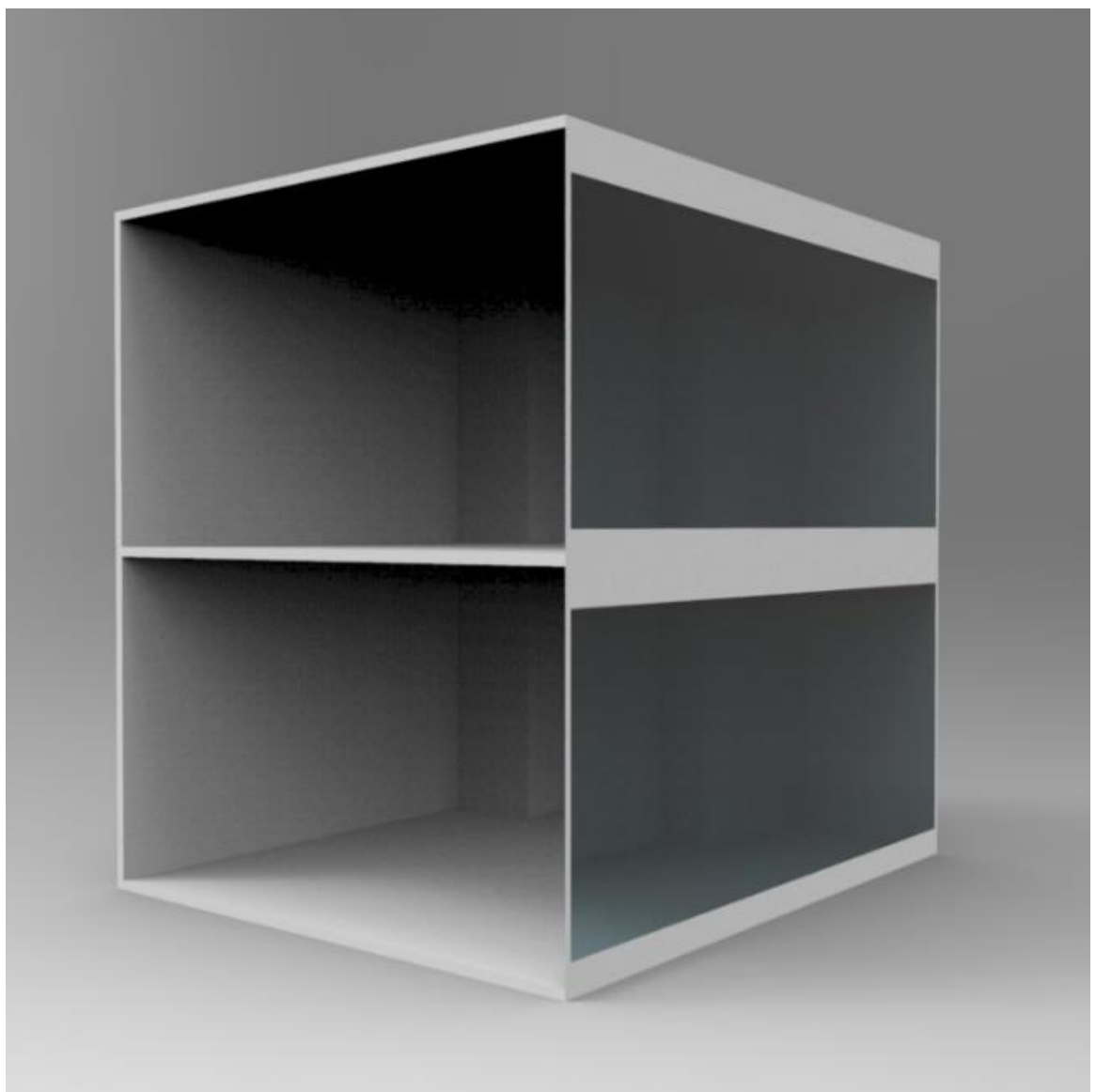

**Figure 1 : 3D view of the studied building**

**Table 1: Basic parameters of the studied building**

| row | Building parameters | specifications |
|---|---|---|
| 1 | Number of floors | Two floors above the ground |
| 2 | floors height | 4 m |
| 3 | floor area | 39.66 $m^2$ |
| 4 | North and South length of the building | 7 m |
| 5 | east and west width | 6 m |
| 6 | Façade direction | South |
| 7 | External walls | ASHRAE 189.1-2009 EXTWALL MASS CLIMATE ZONE 3 |
| 8 | Type of roof | ASHRAE 189.1-2009 EXTROOF IEAD CLIMATE ZONE 2-5 |
| 9 | Type of floors | Interior Floor |
| 10 | types of windows | 7mm+5mm+1.5mm cornice wall |
| 11 | air conditioning system | Ideal Air Loads |
| 12 | operational program | from 8 am to 4 pm |
| 13 | weather data | Shahid Dastghib Airport |

### 2-3- Facade Design Process 2-3-1

### Facade Structure

The facade being analyzed comprises of two Skin, both located on the southern side of the building. The primary Skin includes curtain wall openings that account for 80% of the facade, while the second Skin consists of sixteen modules. Each module consists of 16 elements that can be folded into three separate types, enabling them to create different shapes based on the angle of sunlight. (See Figure 2). An illustration of the kinematic changes of the second Skin is shown below**.**

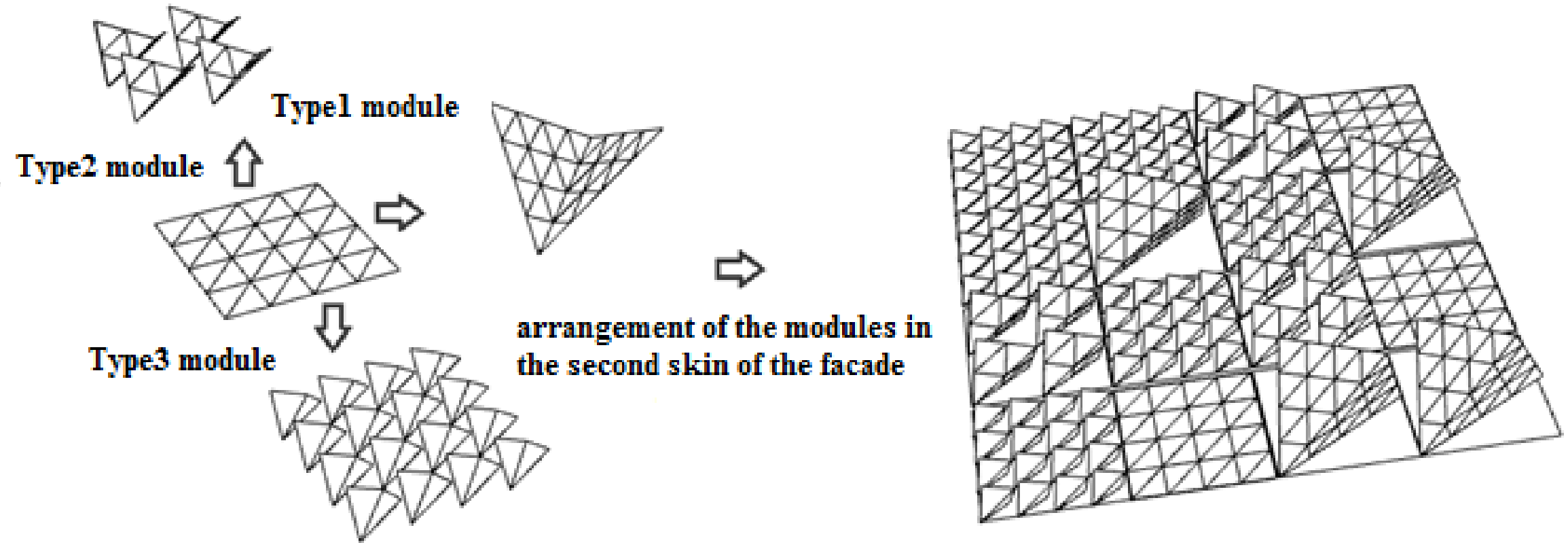

**Figure 2 : The details of the façade structure.**

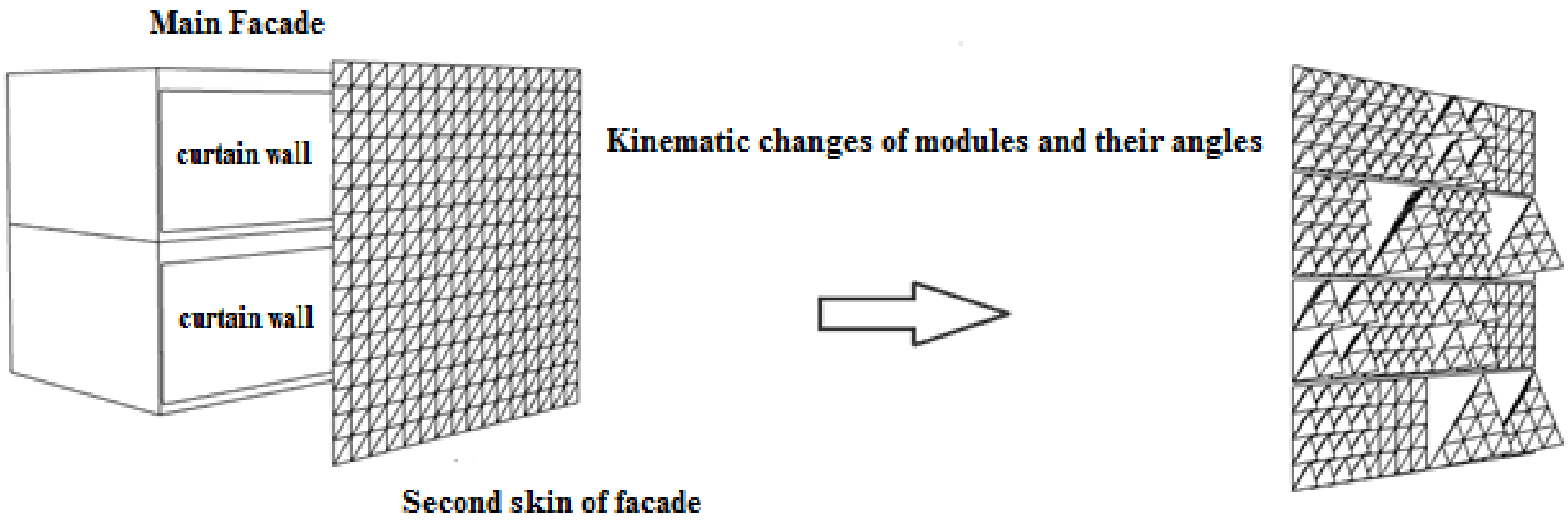

**Figure 3 : The kinematic changes of the facade based on the angle and direction of sun radiation.**

The angle of rotation and the arrangement of the modules' elements are the primary parameters that dictate the kinetic behavior of the facade. The modules are capable of being activated and transformed into various types, with an angular variation range of 0 to 1.57 radians around their main axis. The high degree of freedom in the modules offers immense flexibility to the facade's kinematics. The second skin of the facade is composed of 16 modules, each of which can have up to 16 square-shaped elements. These elements have the ability to alter their angle and fold around their primary axis, contributing to the facade's dynamic characteristics.

**Table 2: Specifications of facade modules**

| Module type | The number of elements in each module | the fold angle matrix of the elements |
|---|---|---|
| 1 | 1 | $[R]$ |
| 2 | 4 | $\begin{bmatrix} R1 & R3 \\ R2 & R4 \end{bmatrix}$ |
| 3 | 16 | $\begin{bmatrix} R1 & \cdots & R13 \\ \vdots & \ddots & \vdots \\ R4 & \cdots & R16 \end{bmatrix}$ |

## 2-4 Optimizing view parameters

### 2-4-1- Big Bang-Big Crunch (BB-BC) algorithm

The selection of appropriate optimization techniques and tools is crucial to overcome inadequacies in optimization [10]. One such technique is the Big Bang–Big Crunch (BB–BC) algorithm, which is based on the theories of the evolution of the universe, namely Big Bang and Big Crunch theory. In the Big Bang phase, energy loss leads to disorder, and randomness is the main characteristic. In contrast, in the Big Crunch phase, randomly distributed particles are absorbed [11]. The BB-BC algorithm has been used in various studies, and Rubik has presented its effectiveness in the passive design of buildings [12].

To achieve a logical answer at the appropriate time and based on several parameters in this research, the BB-BC algorithm was chosen as the appropriate optimization tool. The algorithm starts with the Big Bang phase, which generates random solutions in the search space. In the big crunch phase, the center of mass is calculated based on the inverse of the proportionality function value, which is defined by equation 1.

(equation 1)
$$X_c=\frac{\sum_{i=1}^{n}\frac{1}{f_i}x_i}{\sum_{i=1}^{n}\frac{1}{f_i}}$$

Where $x_i$ is a point in the search space, $f_i$ is the corresponding fitness function value, and n is the population size. In the next stage of the Big Bang, new solutions are generated around the center of mass according to the following equation:

(equation 2)
$$X_i^{k+1}=X_c^k+r_j\alpha\frac{X_{max}-X_{min}}{k+1},\ i=1,\dots,N$$

Where k is the number of iterations, $r_j$ is a random number, and α is a parameter to limit the size of the search space. $X_{max}$ and $X_{min}$ are the maximum and minimum range of solution values. Each time a new population is created in the Big Bang phase, a new center of mass is calculated in the Big Crunch phase. This sequence continues until a criterion for achieving the desired global solution is reached.

## 2-5 Study Stages

### 2-5-1 First Stage

One In the first phase of this study, the building located in Shiraz city is modeled using Grasshopper software and Honeybee and Ladybug plugins based on the specifications provided in Table 1 (Figure 4). After simulating the weather conditions in the period between 1st to 3rd of August, the total operational energy consumption (OEC) of the building is calculated for this time period. In this study, OEC refers to the energy required for air conditioning the building.

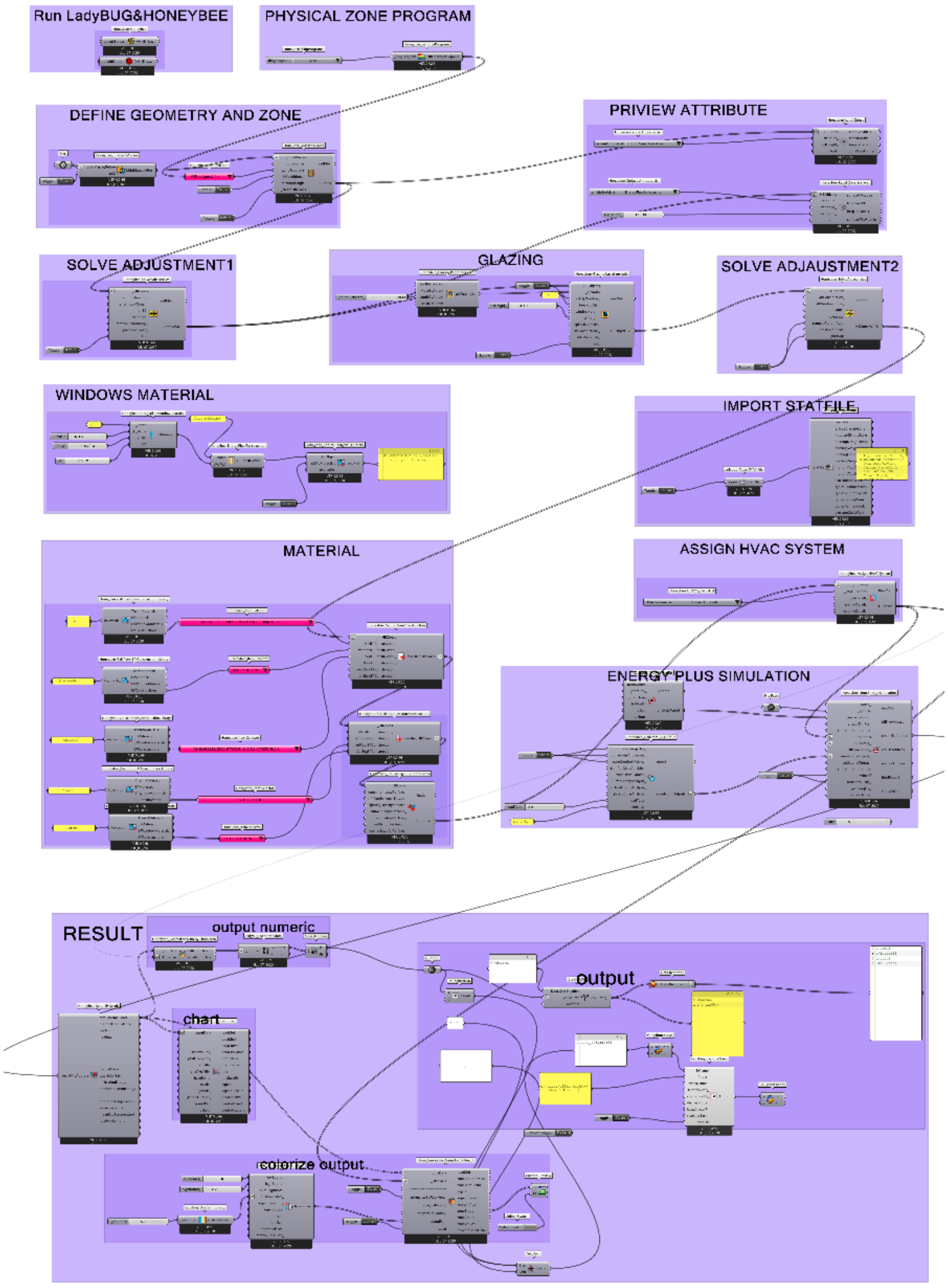

**Figure 4: Simulation process of the model in the first stage using Grasshopper software.**

### 2-5-2 Second Stage

In the second stage, the second façade is parametrically created in Python programming language and Grasshopper software based on the desired parameters and variables. The parameters for the form are optimized using the Big Bang-Big Crunch algorithm to achieve the minimum energy consumption of the building in a three-day period. The best form is selected and its results are examined (Figure 5).

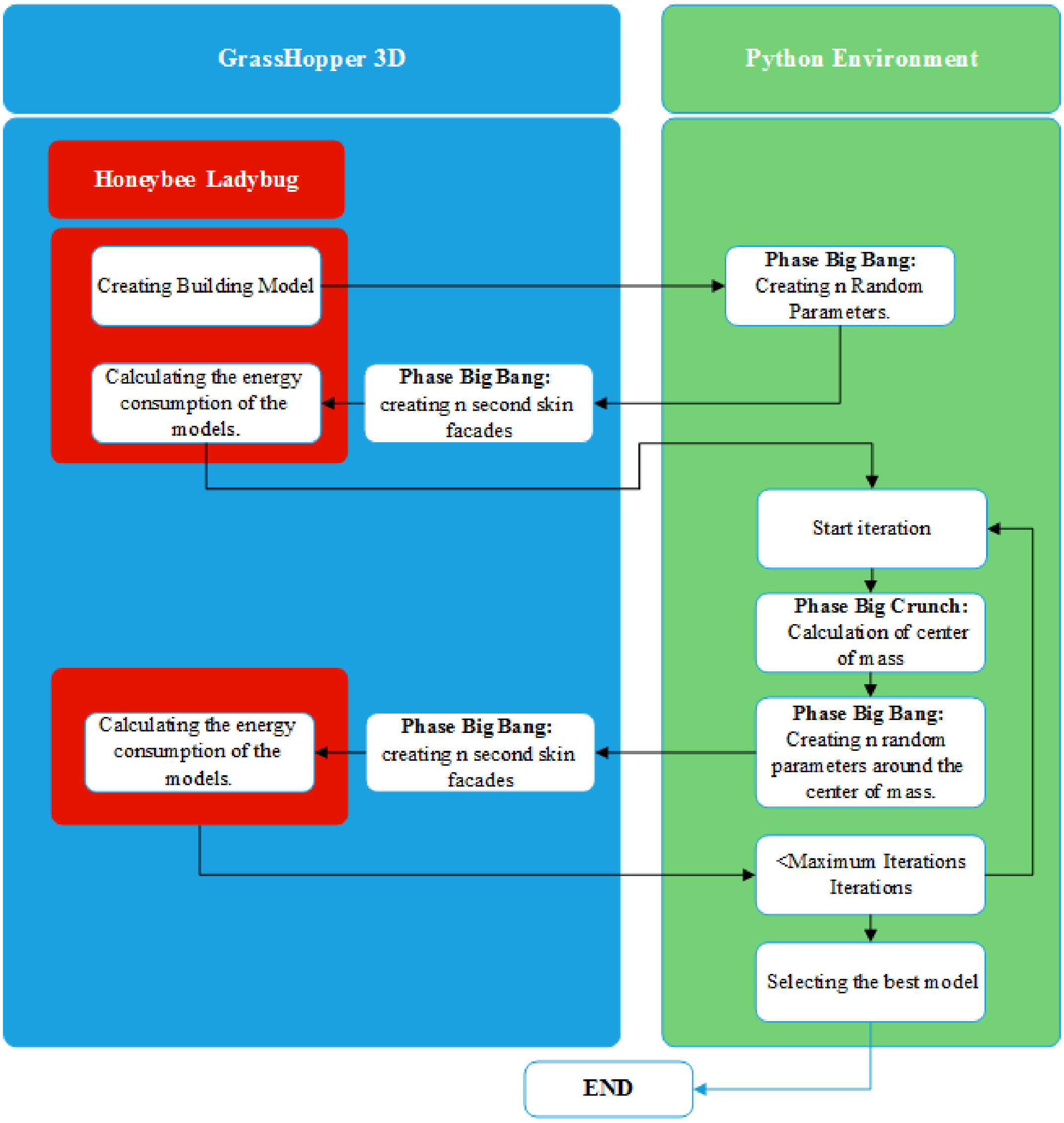

**Figure 5: Framework for creating and optimizing the façade parameters based on the BB-BC algorithm.**

### 2-5-3 third stage

In the third stage, assuming that the second façade is movable and capable of changing its form during a specific period of time in response to the kinetic changes of light and weather, and assuming that the façade's changes can be done in each day of the studied three-day period, the parameters are optimized separately for each time interval using the Big Bang- Big Crunch algorithm. The total energy consumption of these three optimized forms is calculated as the final energy consumption according to the following equation and is studied further.

(equation 3) $$OEC=\sum_{i=1}^{3} OEC_i$$

Where $OEC_i$ represents the energy consumption of the building's operation phase on each day, and OEC is the total energy consumption during the three-day operation phase under onsideration.

## 3- Results and conclusion

In this section, the output data resulting from the analysis of the model are examined to investigate the effects of different stages of modeling on the energy consumption due to operation in the desired time interval.

### 3-1. Results of Stage 1

In the first stage of the study, the energy consumption of the building during the operation phase was investigated in the absence of the second skin facade. The results showed that the energy consumption during the three-day period was 2.76 kWh/m², indicating a relatively high level of energy consumption. The findings of this phase can serve as a baseline for comparing the energy consumption in subsequent stages, where different design strategies are implemented to improve the energy efficiency of the building. (Figure 6)

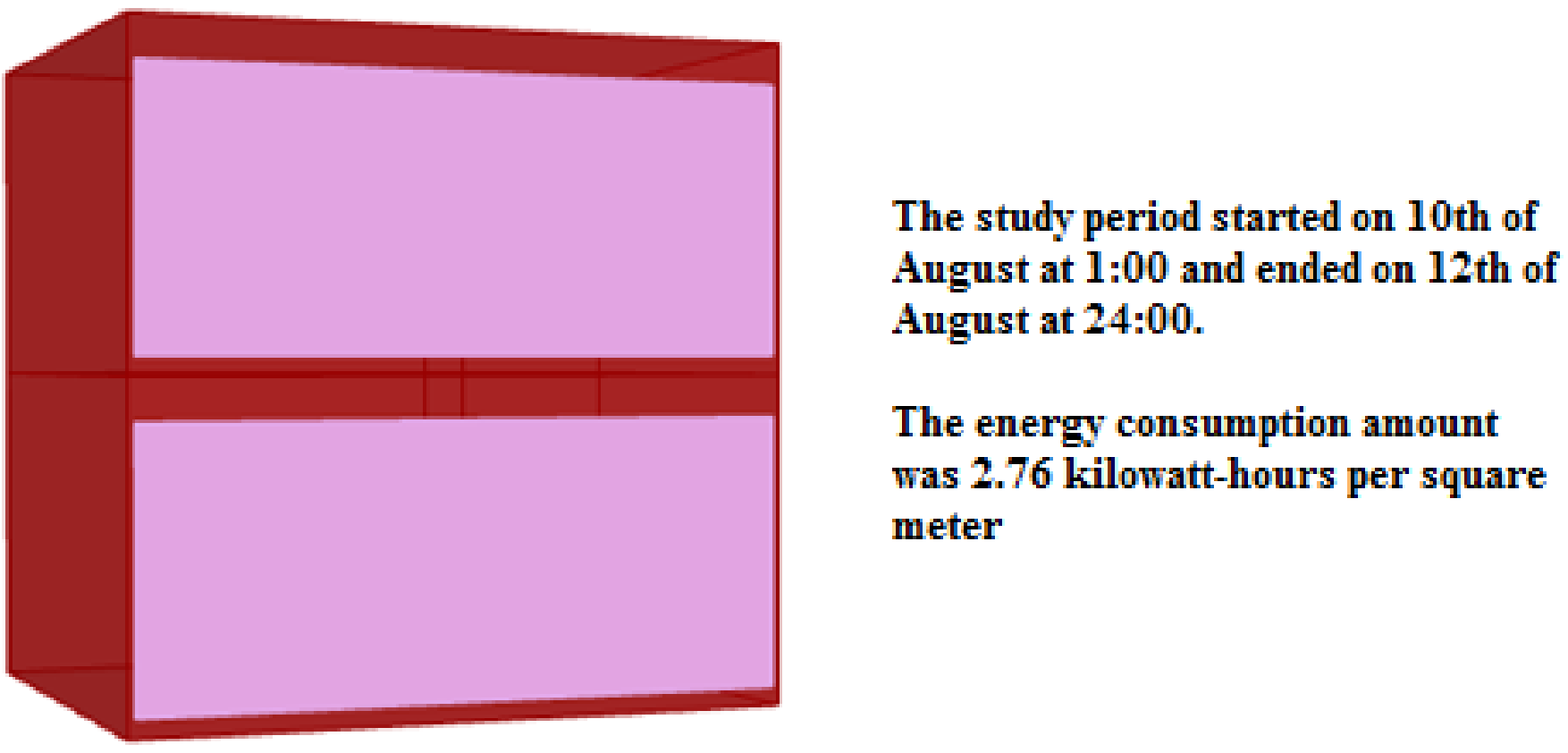

**Figure 6: Energy consumption modeling and results in the first stage of the building**

### 3-2- Results of the Second Stage

In this stage, assuming the static form of the façade, element parameters were optimized using an algorithm and various models were created for the desired time interval. The energy consumption was then optimized to obtain the best solution, Based on the results of the second phase, it can be concluded that by optimizing the parameters of the building envelope elements using an algorithm and creating different models for the desired time period, the energy consumption during the operation phase can be significantly reduced. The optimized energy consumption in the best solution was found to be 2.02 kWh/m2, which represents a 28% reduction compared to the energy consumption in the first phase of the study. These results demonstrate the potential of using computational optimization algorithms to improve the energy efficiency of building design and operation. Table 3, provides details and specifications used in the optimization process.

**Table 3: Details and Specifications Used in the Optimization Process.**

| | | | |
|---|---|---|---|
| Start time of the time interval: | August 1st at 1:00 | End time of the time interval: | August 12th at 24:00 |
| Minimum angle value per element with facade surface: | Rmin=0.50 rad | Maximum angle value per element with facade surface: | Rmax=1.57 rad |
| Number of models created in each iteration: | Sample=3 | Number of iterations in the optimization process: | Iter=15 |
| Optimized energy onsumption after optimization: | OEC=2.02 kwh/$m^2$ | Algorithm parameter: | α =0.3 |

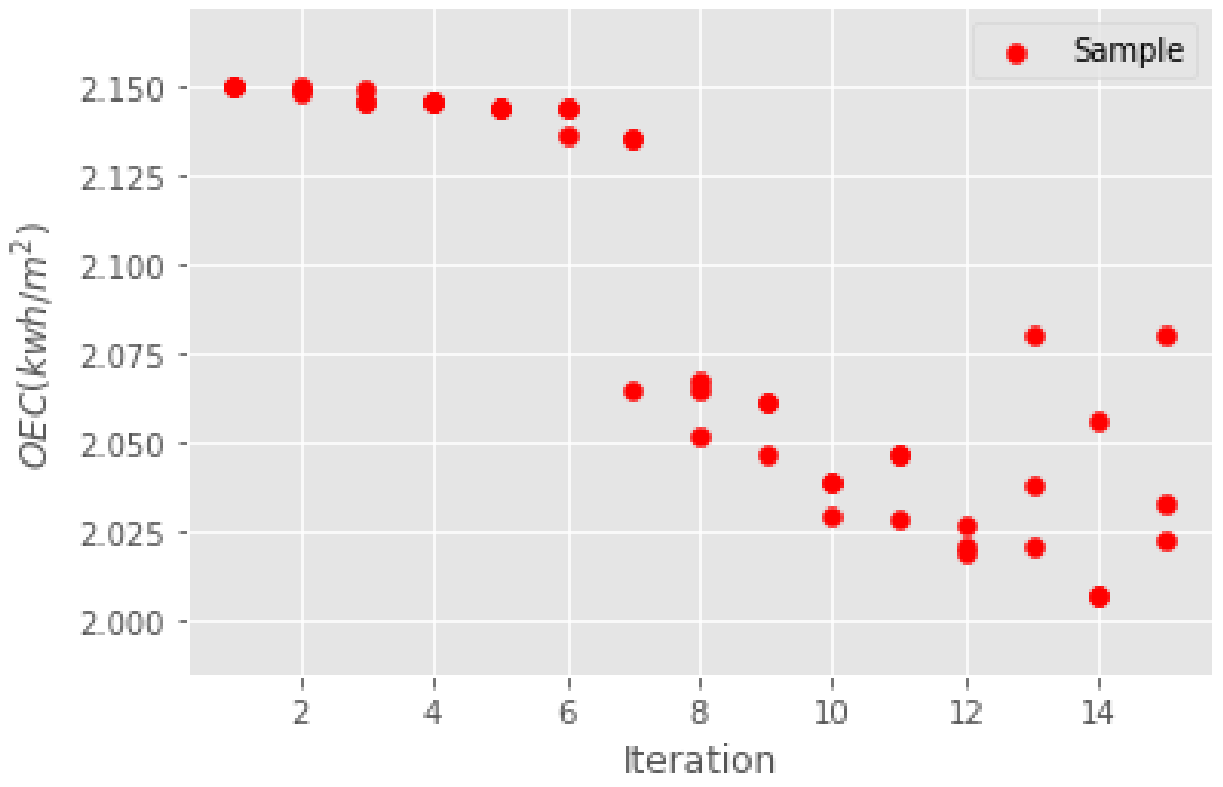

**Figure 7: Optimization process graph of the second façade in phase 2.**

### 3-3- Results of the third Stage

In the third stage, assuming daily changes in parameters and creating a kinematic form, the second facade skin was optimized in three one-day intervals. The details and results of this optimization process are shown in Table 3 and Figure 8.The results of the third phase of optimization for the kinetic façade show a significant improvement in energy consumption. By considering the daily changes in parameters and creating a kinematic form for the façade, the energy consumption has been optimized for each one-day interval. The best results were achieved with an OEC of 0.630 kWh/m2, 0.652 kWh/m2, and 0.658 kWh/m2 for each one-day interval respectively.

**Table 4: Details of the results used in the optimization process in the third phase**

| | | | |
|---|---|---|---|
| Start time of time interval: | August 10, 1:00 | End time of time interval: | August 10, 24:00 |
| Best result after optimization: | OEC=0.630 kwh/m$^2$ | | |
| Start time of time interval: | August 11, 1:00 | End time of time interval: | August 11, 24:00 |
| Best result after optimization: | OEC=0.652 kwh/m$^2$ | | |
| Start time of time interval: | August 12, 1:00 | End time of time interval: | August 12, 24:00 |
| Best result after optimization: | OEC=0.658 kwh/m$^2$ | | |

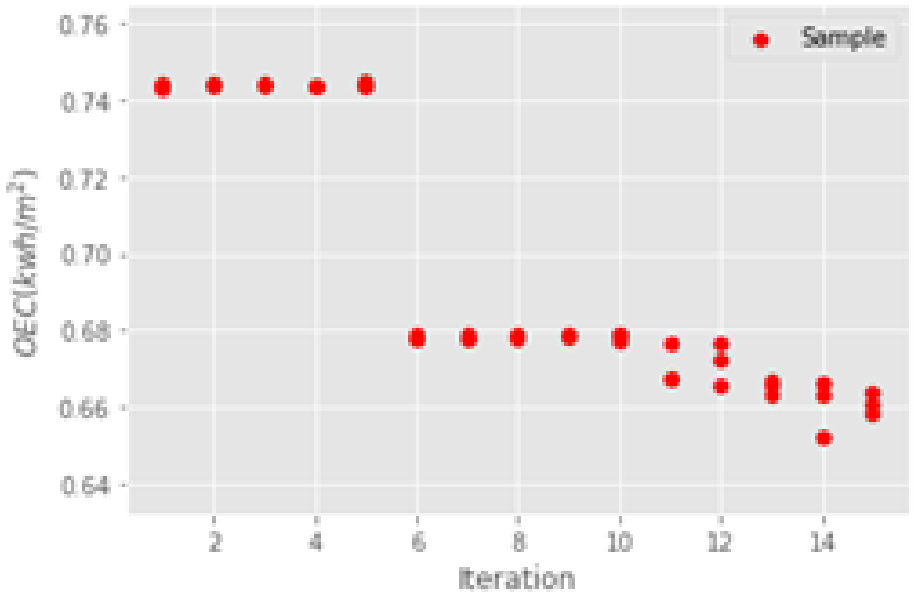

**Optimization process graph of first day**

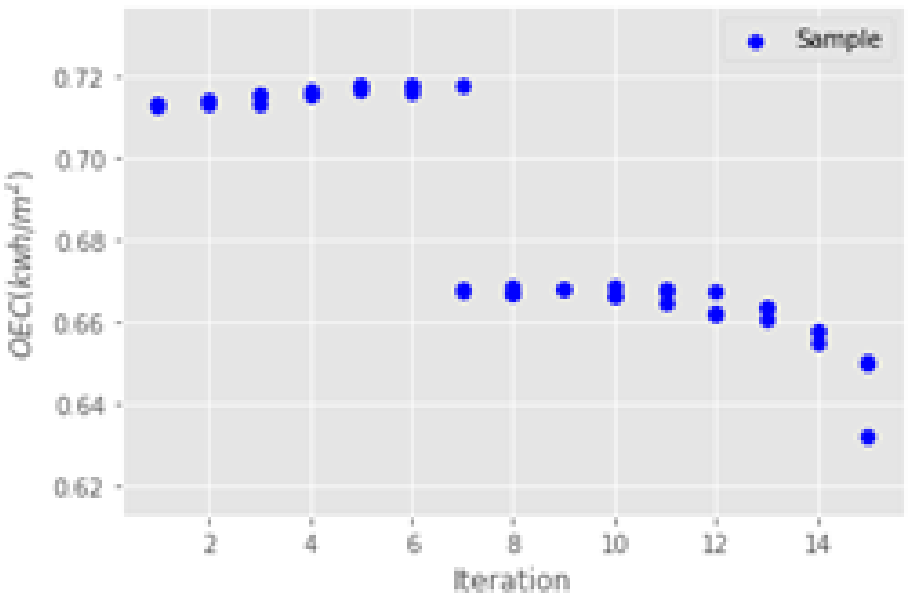

**Optimization process graph of second day**

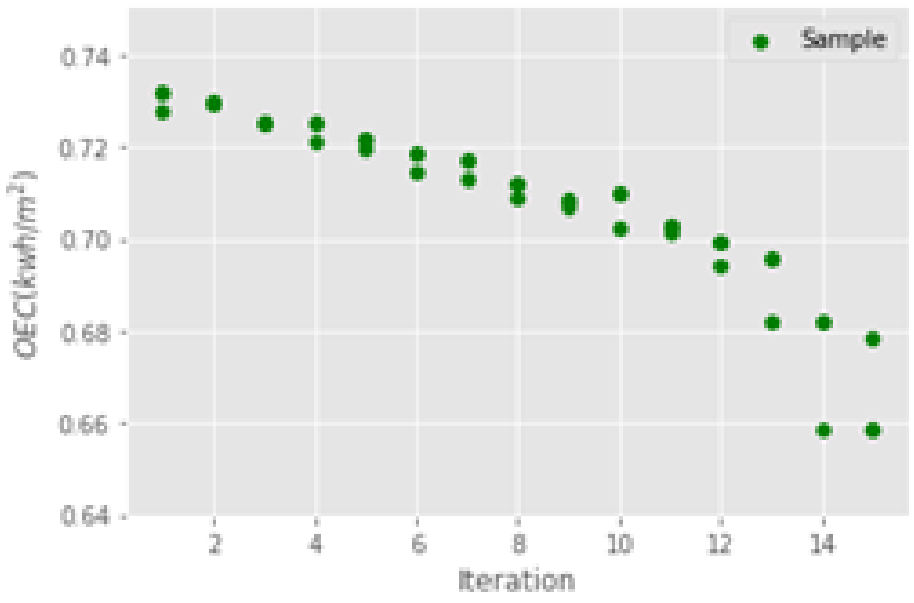

**Optimization process graph of third day**

**Figure 8: Optimization process graph of the second façade in phase 3.**

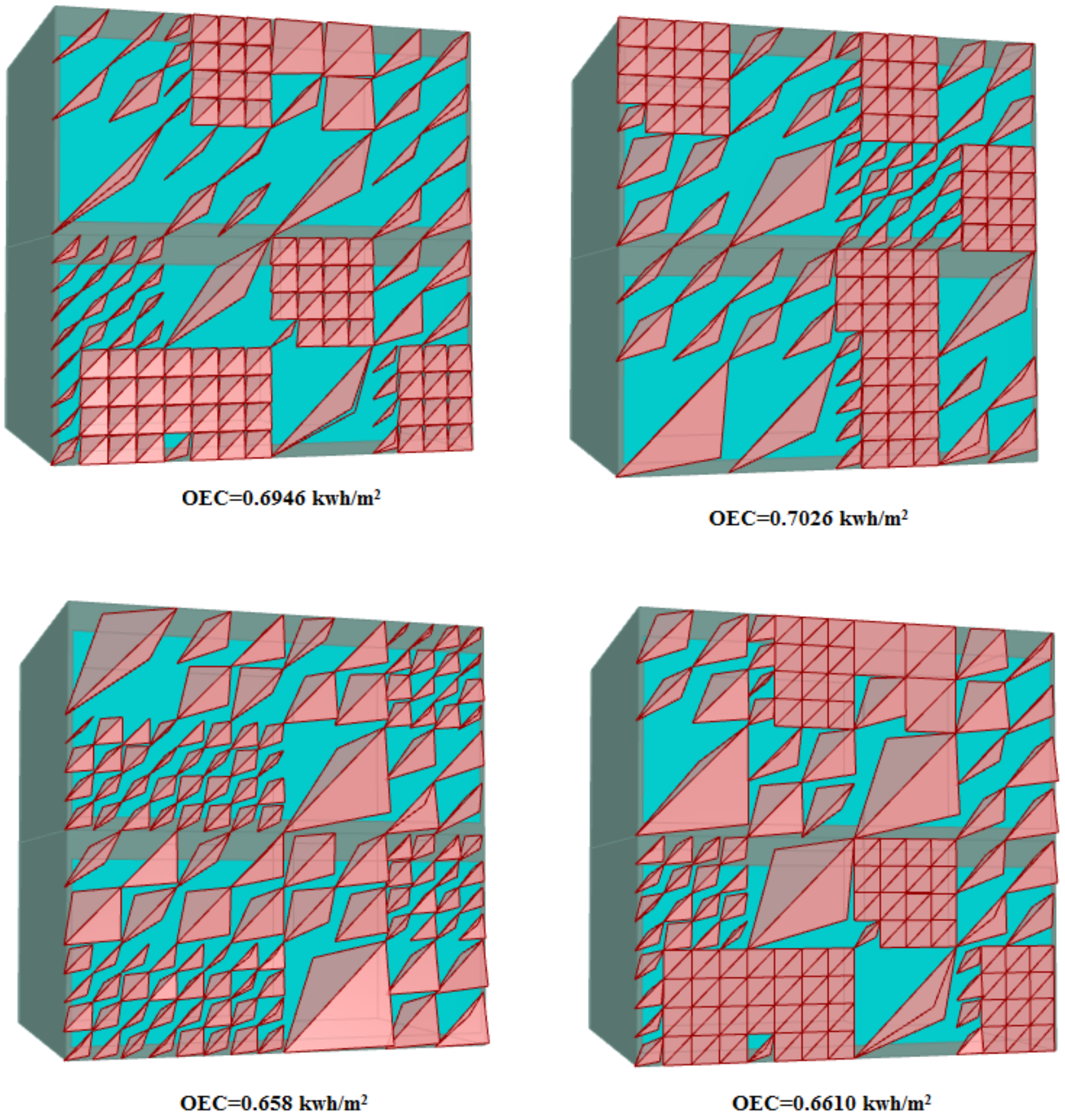

**Figure 9: Model changes in the optimization process of the second façade**

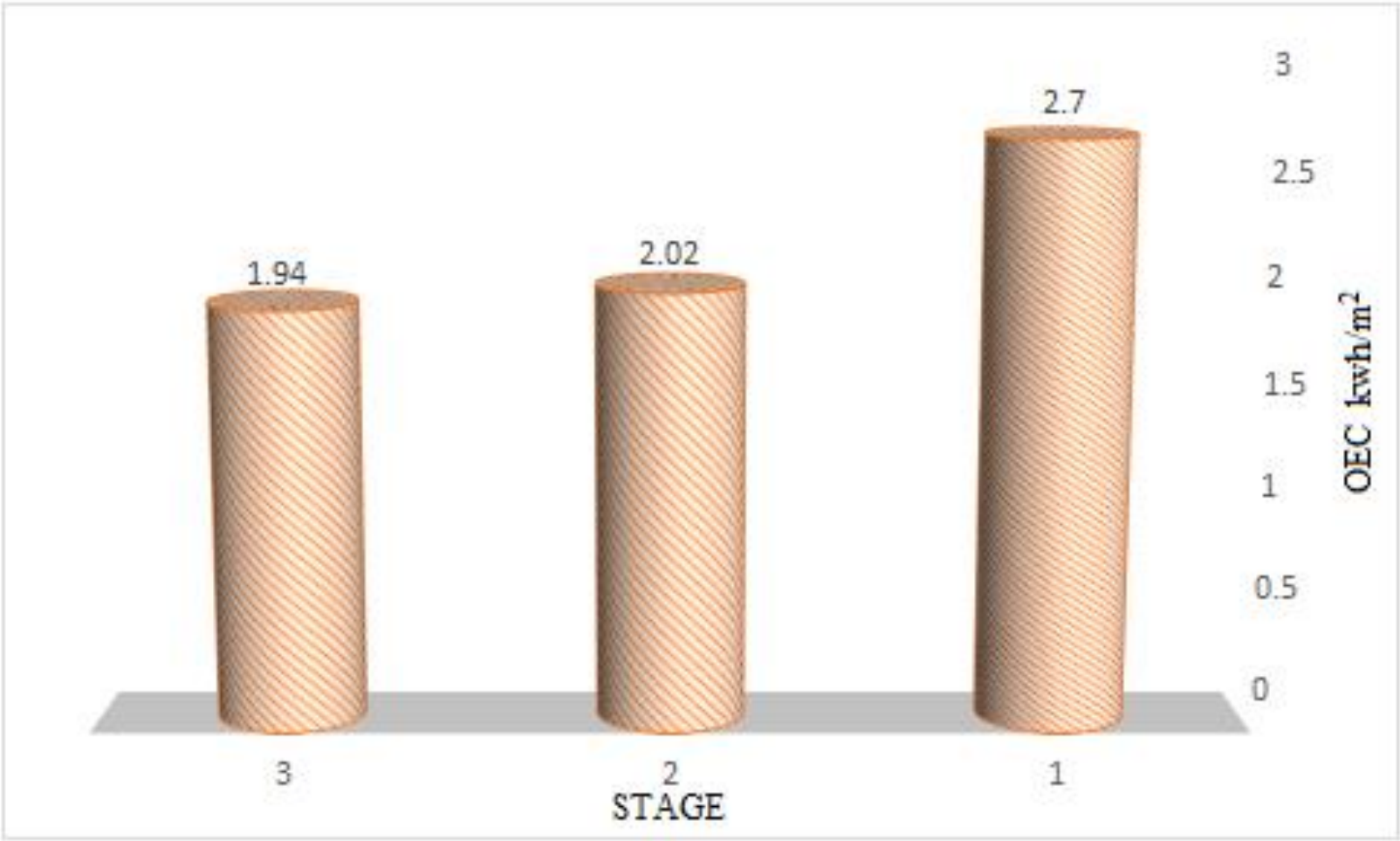

**Figure 10: Comparison of energy consumption in three stages of the study**.

## 4- Conclusion

The purpose of this study was to propose a kinetic facade and investigate its performance in an office building. The comparison of graphs (Figure 10) indicates that the use of a parametric double-skin facade with an optimized form has resulted in a 25% reduction in energy consumption in the building, while the use of a parametric double-skin facade with kinetic performance has led to a 28% reduction in energy consumption. Based on the results of the second and third stages of the study, the use of a facade with kinetic performance has resulted in a 4% reduction in energy consumption compared to its static performance. Considering the limited number of days studied and the proximity of the studied days, the result is acceptable. The results of this study can be complemented by a multi-objective study that includes not only reducing energy consumption but also providing daylight. Moreover, increasing the number of days studied and reducing the time interval of changing the parameters of the kinetic facade can provide more realistic results and help in the development of an intelligent kinetic system. In conclusion, the results of this study indicate that the use of a facade with kinetic performance can significantly reduce energy consumption in buildings. Therefore, it is recommended that architects and designers consider this technology in their designs to create more energy-efficient buildings.

## 5-Refrences